\documentclass[prd,preprint,showpacs,showkeys,preprintnumbers,amsmath,amssymb,floatfix,nofootinbib]{revtex4}

\usepackage[utf8]{inputenc}
\usepackage{hyperref}
\usepackage{verbatim}
\usepackage{dcolumn}
\usepackage{bm}
\usepackage{graphicx}
\usepackage{color}
\usepackage{subfigure}

\def\be{\begin{equation}}
\def\ee{\end{equation}}
\def\bestar{\begin{equation*}}
\def\eestar{\end{equation*}}

\DeclareMathOperator{\arccot}{arccot}
\DeclareMathOperator{\arcsinh}{arcsinh}
\DeclareMathOperator{\arccoth}{arccoth}
\DeclareMathOperator{\arctanh}{arctanh}
\DeclareMathOperator{\csch}{csch}

\begin{document}

\title{Nonlinear extensions of gravitating dyons: from NUT wormholes to Taub-Bolt instantons}

\author{Daniel Flores-Alfonso}
\email{daniel.flores@cinvestav.mx}
\affiliation{Departamento de F\'{i}sica, CINVESTAV-IPN, 
                  A.P. 14-740, C.P. 07000, Ciudad de México, Mexico}
                  
\author{Román Linares}
\email{lirr@xanum.uam.mx}
\affiliation{Departamento de F\'{i}sica \\ Universidad Aut\'{o}noma Metropolitana - Iztapalapa\\
                   Av. San Rafael Atlixco 186, C.P. 09340, México City, M\'exico}         
                   
\author{Marco Maceda}
\email{mmac@xanum.uam.mx}
\affiliation{Departamento de F\'{i}sica \\ Universidad Aut\'{o}noma Metropolitana - Iztapalapa\\
                   Av. San Rafael Atlixco 186, C.P. 09340, México City, M\'exico}

\date{\today}

\begin{abstract}

Recent work has shown the existence of a unique nonlinear extension of electromagnetism which preserves conformal symmetry and allows for the freedom of duality rotations. Moreover, black holes and gravitational waves have been found to exist in this nonlinearly extended electrovacuum. We generalise these dyonic black holes in two major ways: with the relaxation of their horizon topology and with the inclusion of magnetic mass.
Motivated by recent attention to traversable wormholes, we use this new family of Taub-NUT spaces to construct AdS wormholes. We explore some thermodynamic features by using a semi-classical approach. Our results show that a phase transition between the nut and bolt configurations arises in a similar way to the Maxwellian case.
\end{abstract}



\maketitle

\section{Introduction and Motivation}

The Taub-NUT (Newman, Unti, Tamburino) geometry is one of the most captivating, and at times puzzling, vacuum solutions of General Relativity. 
The interior Taub-NUT spacetime was initially found as a particular Bianchi IX cosmology~\cite{Taub:1950ez}, one which is homogeneous but anisotropic. Thus, the $su(2)$ isometry algebra naturally led to a hyperspherical topology on spatial slices. This cosmological metric specialises to that of Kantowski-Sachs~\cite{Kantowski:1966te} which is locally equivalent to the interior Schwarzschild solution. This result matches perfectly with the fact that the exterior Taub-NUT metric~\cite{Newman:1963yy} is a one-parameter generalisation of the Schwarzschild black hole. Notwithstanding, the iconic Schwarzschild geometry is spherically symmetric, leading to a distinct preferred topology. This conflict of spacetime topology is but a mere taste of the sort of perplexities that NUTty configurations have to offer.

In this paper, we present a family of Taub-NUT spacetimes sourced by electromagnetic matter complying with a nonlinear constitutive relation. Our motivation is multipronged. Let us start by mentioning that the sourcing field theory is conformal and its motion is invariant under electric-magnetic duality rotations, which are precisely the symmetries of Maxwell theory. Imposing them on an arbitrary nonlinear electrodynamics yields a unique nonlinear extension of Maxwellian dynamics, henceforth referred to as ModMax; this theory has been very recently found in~\cite{Bandos:2020jsw} and studied in~\cite{Kosyakov:2020wxv2}.

Fairly recent studies on traversable Taub-NUT wormholes~\cite{Ayon-Beato:2015eca,Clement:2015aka} have conducted to developments in chronology protection, gravitational symmetry breaking, spacetime thermodynamics and finite Baryon density phase structure~\cite{Clement:2016mll,Canfora:2017ivv,Kubiznak:2019yiu,Ayon-Beato:2019tvu}. These results have driven us to study wormholes of a NUTty nature see, e.g., Ref.~\cite{Anabalon:2018rzq,Anabalon:2020loe} which have applications in holography. A key feature of these particular configurations is that they possess hyperbolic symmetry\footnote{As opposed to spherical symmetry which is the root of timelike curves in NUT-like spacetimes.}. Here, we present a generalisation of these hyperbolic NUT wormholes whilst examining the effects of allowing a nonlinear constitutive relation for the electromagnetic sector. For a recent example of a nonlinearly charged wormhole, we refer the reader to~\cite{Canate:2020btq}. Let us recall, once more, that the particularity of the electrodynamic extension chosen in this paper is that we preserve the maximum allowable amount of symmetries of the Maxwell equations.

Our investigation is not only motivated by traversable wormholes but also by nonlinear electrodynamics in general. Nonlinear constitutive relations in electrodynamics conduct to fascinating theoretical models and effectively account for non-trivial phenomena~\cite{Born:1934gh,Heisenberg:1935qt}. Born-Infeld theory is a case in point; it is the unique regular nonlinear electrodynamics that does not present shockwaves or birefringence~\cite{BialynickiBirula:1984tx}. Another significant example is Euler-Heisenberg theory which describes photon-photon scattering in quantum electrodynamics. Furthermore, when nonlinear electrodynamics is coupled to Einstein gravity, a plethora of modifications to electrovacuum configurations arise. To name a few: 1) Geodesics on Born-Infeld black holes~\cite{GarciaD2007} may reach the spacetime singularity, as opposed to their Maxwellian counterparts~\cite{Breton:2002td,Linares:2014nda}; this trait is more akin to geodesics in Schwarzschild geometry. 2) It is commonplace for the nonlinearity parameters to act as screening factors for the field charges, see, e.g.,~\cite{Ruffini:2013hia,Maceda:2018zim,Flores-Alfonso:2020euz}. 3) Nonlinear constitutive relations led to the first regular black hole configurations, see~\cite{AyonBeato:1998ub} and references therein. In this respect, gravity together with nonlinear electrodynamics, have led to surprising results. What is more, nonlinearly charged black holes have been studied in the context of holographic superconductors~\cite{Sheykhi:2015mxb,Maceda:2019woa}.
Motivated by the novel matter content of ModMax non-linear electrodynamics, in this work we expand on earlier findings~\cite{Flores-Alfonso:2020euz} with a major interest in the thermodynamic structure; in spite of the non-linearity of the electrodynamics, the analysis of the phase transition between the nut and bolt configurations can be done analytically up to a certain extent. 

Now, let us recall some well-known results on Taub-NUT spacetime. First, in terms of Schwarzschild-like coordinates, the interior Taub-NUT metric may be written as
\begin{equation}
 ds^2=-U(r)^{-1}dr^2+U(r)(dt+2n\cos\theta d\phi)^2
 +(r^2+n^2)(d\theta^2+\sin^2\theta d\phi^2), 
 \label{Taub}
\end{equation}
with the metric function $U$ given by
\begin{equation}
 U(r)=(-r^2+2mr+n^2)/(r^2+n^2).
\end{equation}
It is straightforward to see that $\xi=\partial/\partial t$ is a Killing vector field. Moreover, the roots of $U$ indicate the presence of two Killing horizons. In the black hole limit, $n\to0$, one of the horizons transmutes into the Schwarzschild curvature singularity, whilst the other becomes the event horizon. For the black hole, the locus of $\theta=0,\pi$ is a mere coordinate degeneracy; this is not the case for the Taub-NUT metric, where the locus has been termed Misner string~\cite{Misner:1963fr}. However, if one chooses that spacetime has $R\times S^3$ topology, then the spherical atlas on $S^3$ renders the Misner string unobservable. The spacetime is extendable through the horizon via Eddington-Finkelstein coordinates. There are two isometric extensions, $M_{\pm}$ say. If this is carried out on both horizons, we have four combinations but two of them are isometric, and the remaining are not isometric among themselves, say $M_{++}$ and $M_{-+}$~\cite{Chrusciel:1993yy}. The origin of having different extensions through each horizon is that neither one of the extension $M_{+}$ or $M_{-}$ provides a complete treatment of the horizons. The nature of the Eddington-Finkelstein transformation makes geodesics approaching one horizon branch pass through normally, whilst those approaching the other branch spiral indefinitely~\cite{Misner:1968}. A significant price to pay is that these maximal extensions possess closed timelike curves.

An alternative to the maximal extension presented above is available and is based on Kruskal-Szekeres coordinates~\cite{Miller:1971em}. This extension is incompatible with the previous one, e.g., if one adopts both extensions, then spacetime loses its Hausdorff property.  Although absent of curvature singularities this extension would in principle be geodesically incomplete, due to the Misner string, perhaps not surprising in the eyes of the singularity theorems. Nonetheless, recent studies on this extension have shown that the Misner string is completely transparent to the geodesics, since, geodesics do not stop there. From this point of view, this extension is geodesically complete~\cite{Clement:2015aka,Clement:2015cxa}. For full analytic details on geodesics (complete and incomplete) for Eddington-Finkelstein and Kruskal-Szekeres extensions, we refer the reader to~\cite{Kagramanova:2010bk}, whereas for the (nonlinearly) charged case we suggest~\cite{Breton:2014mba}.

To help us interpret the NUT parameter it has been enlightening to compare gravity with electromagnetism~\cite{Ellis:1971pg}.
Let us start by mentioning that, as an Einstein manifold, all the spacetime curvature of Taub-NUT is encoded into the Weyl tensor. The electric part of the tensor is sourced by the mass giving rise to the interpretation of mass as the gravitoelectric charge. The magnetic part of the Weyl tensor is entirely analogous but with the NUT parameter corresponding to the gravitomagnetic monopole moment\footnote{When this analysis is carried out for the Kerr metric, one finds that angular momentum is the gravitomagnetic dipole moment.}. Results such as these have earned the NUT parameter the term magnetic mass. This analogy between the Weyl tensor and the Maxwell field strength is far from being the only one.

Electric-magnetic duality invariance is a signature
symmetry of Maxwell's equations. In a recent work~\cite{Huang:2019cja}, electric-magnetic duality transformations have been shown to correspond to the long-established complex transformations~\cite{Talbot:1969bpa,Quevedo:1992} relating the Schwarzschild metric to the Taub-NUT geometry. As it turns out, at null infinity these transformations are equivalent to complexified Bondi-Metzner-Sachs (BMS) supertranslations. For earlier works on dualities, analogies between electric-magnetic charges and masses, as well as, how they relate to angular momentum we recommend~\cite{PhysRevD.17.1957, Garfinkle:1990zx,NouriZonoz:1998cp,Turakulov:2001bm,Turakulov:2001jc,Bunster:2007sn}.

As previously mentioned, both the NUT parameter and angular momentum source the magnetic part of the Weyl
tensor. This, at the very least, suggests that the two quantities are related somehow. Indeed, the approach taken in~\cite{Bonnor:1969}, and more recently in~\cite{Manko:2005nm}, points in this direction. Therein, the Misner string is considered a source of angular momentum, e.g., a topological defect of sorts. In an attempt to clarify this interpretation, we mention that the NUT sector of the metric \eqref{Taub}, where the Killing field $\xi$ is timelike, has a non-diagonal term which causes a Lense-Thirring effect. However, most notoriously, frame dragging in the northern hemisphere is in the opposite direction as in the southern hemisphere. In this sense, the nonstatic nature of the Taub-NUT metric is related to angular velocity but distributed in such a way that total angular momentum vanishes; for a recent discussion see~\cite{Durka:2019ajz}.
When a negative cosmological constant is considered, the Ashtekar-Magnon-Das method of conformal completion~\cite{Ashtekar:1984zz,Ashtekar:1999jx} allows for interesting definitions of mass and angular momentum. These charge definitions reveal that the NUT parameter is related to both these quantities. Specifically, Misner strings source angular momentum~\cite{Kubiznak:2019yiu}. In a complementary manner, in Euclidean configurations, adding topological terms to the action shows that the NUT parameter contributes to the mass~\cite{Araneda:2016iiy}. Moreover, from a thermodynamic point of view Misner strings are sources of entropy\cite{Hawking:1998jf,Ciambelli:2020qny}. Lastly, we mention that from an astrophysical point of view, NUT parameters have observable effects on the apparent horizon's shadow~\cite{Zhang2021}. They also
create a distinctive type of gravitational lensing which may, in principle, be observed~\cite{LyndenBell:1996xj,NouriZonoz:1998va}.

This paper is organised as follows: in Sec.~\ref{secc:1} we review the nonlinear electrodynamics recently proposed by Bandos et. al.~\cite{Bandos:2020jsw}.
We write down the gravitational field equations for the Einstein-ModMax system with cosmological constant and solve them to obtain new spacetime solutions of Taub-NUT type in Sec.~\ref{secc:2}. Afterwards, we present a closely related traversable wormhole configuration with hyperbolic symmetry in Sec.~\ref{secc:2a}. We then continue to Sec.~\ref{secc:2b} where we find Euclidean analogues of the previously mentioned Taub-NUT family.
Then we proceed to discuss in detail the nut and bolt solutions in Sec.~\ref{secc:3}. Using the counter-term subtraction procedure, we analyse how the thermodynamic phase structure and the phase transition between the nut and bolt solutions depend on the non-linear parameter of ModMax electrodynamics; its relation to the Maxwellian case is also discussed. Finally, we comment on our results and perspectives in the Conclusions. We set $c = 1 = G$ throughout the text.

\section{ModMax Theory}
\label{secc:1}

As the electromagnetic theory we consider in this work is quite recent~\cite{Bandos:2020jsw,Kosyakov:2020wxv2}, we describe it in some detail in this section. Before we do, let us recall that Maxwell electrodynamics enjoys a number of symmetries, i.e., invariance under electric/magnetic duality rotations, Lorentz transformations and conformal mappings. These symmetries have a long history in physics and mathematics, for conformal symmetry we refer the reader to~\cite{Bateman:1909,Cunningham:1910,Bateman:1910a,Bateman:1910b,Bessel-Hagen:1921}
while for SO(2) Hodge duality invariance we suggest~\cite{Rainich:1925,Deser:1976,Gaillard:1981,Salazar:1987ap,Gibbons:1995cv,Olive1996}.

Let us start with the general framework of an arbitrary nonlinear electrodynamics. We consider a Lagrangian map $L$ parametrised by two independent quantities, say $x$ and $y$. They constitute the only two independent Lorentz invariant quantities which can be constructed from the Faraday tensor and the metric in four dimensions, i.e.,
\begin{equation}
 x:=\frac{1}{4}F_{\mu\nu}F^{\mu\nu}, \quad\text{and}\quad y:=\frac{1}{4}F_{\mu\nu}\star F^{\mu\nu},
\end{equation}
where $\star$ is the Hodge star linear map. The Euler-Lagrange equation is
\begin{equation}
 d\star P=0, \label{dstarP}
\end{equation}
where $P$ is defined by differentiating $L$ with respect to the electromagnetic tensor $F$. Specifically, we mean
\begin{equation}
 P=-L_xF-L_y\star F, \label{constitutive}
\end{equation}
which is the theory's constitutive relation. 
Note that we are using $L_x$ to indicate the derivative of $L$ with respect to $x$, and similarly for $y$.
On the other hand, one has the energy momentum tensor $T$, which has components given by
\begin{equation}
 4\pi T^\mu{}_\nu = F^{\mu\alpha} P_{\nu\alpha} +   L\delta^\mu_\nu. \label{Tmunu}
\end{equation}
It is often convenient to work with Pleba\'nski's dual variables~\cite{Plebanski:1970,Gibbons:2001sx}
\begin{equation}
 s:=-\frac{1}{4}P_{\mu\nu}P^{\mu\nu}, \quad\text{and}\quad t:=-\frac{1}{4}P_{\mu\nu}\star P^{\mu\nu}.
\end{equation}
In this case, the Legendre transform $\hat{L}(s,t)$ defined by
\begin{equation}
 \hat{L}=-\frac{1}{2}F_{\mu\nu}P^{\mu\nu}-L
\end{equation}
describes the dynamics of the system. For a recent application of this alternative see~\cite{Flores-Alfonso:2020euz}.

ModMax theory is the unique conformal U(1) gauge theory with a nonlinear constitutive relation which enjoys electric/magnetic duality rotation invariance. By considering an arbitrary
nonlinear electrodynamics, as above, imposing the conditions
\begin{equation}
 T^\mu{}_\mu =0\quad\text{and}\quad P_{\mu\nu}\star P^{\mu\nu}= F_{\mu\nu}\star F^{\mu\nu}, \label{cond}
\end{equation}
yields a system of partial differential equation for $L$ which admits the solution~\cite{Kosyakov:2020wxv2}
\begin{equation}
 L=-x\cosh\gamma+\sinh\gamma\sqrt{x^2+y^2}.
 \label{MML}
\end{equation}
The first of equations \eqref{cond} is the necessary and sufficient condition for $L$ to be invariant under conformal transformations~\cite{Bessel-Hagen:1921}. The second guarantees that the equations of motion are invariant under Hodge transformations~\cite{Gaillard:1981,Salazar:1987ap,Gibbons:1995cv}.

Let us emphasise a key difference between the symmetries discussed above. As in Maxwell theory, conformal invariance is a symmetry of the action and so as a consequence, in 4 dimensions, the trace of the energy momentum tensor vanishes~\cite{Bessel-Hagen:1921}. This contrasts with duality rotations which are not a symmetry of the Maxwell Lagrangian; they are a freedom of the equations of motion. In nonlinear electrodynamics duality rotations may be implemented in \emph{a priori} different ways [see, e.g., Ref.~\cite{Buratti:2019cbm}]. The criterion we undertake was first examined in~\cite{Gaillard:1981} and was further developed in~\cite{Gibbons:1995cv}. It was also independently derived in~\cite{Salazar:1987ap}.

The central idea is to demand invariance of the constitutive equation so that the equations of motion admit duality invariance whilst recovering Maxwell theory in the weak field limit. The necessary and sufficient condition for this to happen is the second one in \eqref{cond}. In a complementary manner, the Hamiltonian approach taken in \cite{Bandos:2020jsw} shows that this theory is the unique nonlinear extension of Maxwell electrodynamics which preserves its invariance under conformal transformations and duality rotations.

To close this section, we mention that the theory's nonlinearity parameter $\gamma$ is dimensionless, courtesy of its conformal invariance. When $\gamma=0$ Maxwell theory is recovered, i.e., $L=-x$. In other words, a small value 
of $\gamma$ corresponds to the linear theory. Moreover, the ModMax nonlinear extension is parity conserving, i.e., $L(x,-y)=L(x,y)$. Additionally, observe that the constitutive relation 
\eqref{constitutive} for ModMax is precisely
\begin{equation}
 P=\left(\cosh\gamma-\frac{\sinh\gamma\, x}{\sqrt{x^2+y^2}}\right)F-\frac{\sinh\gamma\, y}{\sqrt{x^2+y^2}}\star F. \label{CRMM}
\end{equation}

\section{A new family of Taub-NUT spacetimes}
\label{secc:2}

We are interested in the coupled theories of General Relativity and ModMax
\begin{equation}
 I[g,A]=-\frac{1}{16\pi}\int d^4x\sqrt{-g}\left(R-2\Lambda-4L\right), \label{action}
\end{equation}
where $L$ is the ModMax Lagrangian in \eqref{MML}.
The Euclidean version of this action is related to the calculations carried out in the sections \ref{secc:2b} and \ref{secc:3}.

We consider first Taub-NUT geometries
with spherical, hyperbolic or planar symmetry.
They are sourced by ModMax matter and allow for a non-zero cosmological constant. Spacetimes with different symmetry are of interest to us for distinct reasons. For example, hyperbolically symmetric geometries allow for traversable wormholes but do not allow for gravitational instantons with nuts~\cite{Chamblin:1998pz}.

For now, if only for generality, let us consider a metric Ansatz given by
\begin{equation}
ds^2 = -f(r) (dt + 2n{\cal A})^2 + f(r)^{-1} dr^2 +  (r^2 + n^2)d\sigma^2.
\label{taubnut}
\end{equation}
These spacetimes are radial extensions of line bundles over K\"ahler manifolds~\cite{Page:1985bq}. The base manifolds
have line element $d\sigma^2$ and symplectic potential one-form ${\cal A}$. Thus, the space's symplectic form is given by $\omega=d{\cal A}$. We consider three K\"ahler spaces: the Euclidean plane, the Riemann sphere and the Lobachevsky plane. We parametrise these spaces so that they have constant sectional curvature $\kappa=0,+1,-1$, respectively.

In analogy with Einstein-Maxwell~\cite{Brill:1964,Carter:1968,Ruban:1972} and Einstein-Born-Infeld~\cite{GarciaD2007} configurations, we choose fields aligned with the metric. To this end, we consider a gauge potential Ansatz of the form
\begin{equation}
 A := a(r) \ell := \frac 1{2n} [q \sin \Phi(r) + g \cos \Phi(r)] \ell, \label{gauge}
\end{equation}
where we have defined $\ell := dt + 2n{\cal A}$; notice that $d\ell= 2n\omega$. For Maxwell matter, $\Phi$ is given by an inverse trigonometric function whilst for Born-Infeld, it is an inverse elliptic function.
The ModMax field equations, cf. \eqref{dstarP} and \eqref{CRMM}, yield
\begin{equation}
\Phi (r) = 2n e^{-\gamma} \int_r^\infty \frac {dr}{r^2 + n^2} = 2 e^{-\gamma} \left[ \frac \pi2 - \arctan \left( \frac r n \right) \right] = 2 e^{-\gamma} \arccot\left( \frac r n \right).
\label{Phi}
\end{equation}
The Maxwell case is recovered setting $\gamma=0$~\cite{Brill:1964}; this value of $\gamma$ also allows for the simplification of \eqref{gauge} using trigonometric identities and resulting in an algebraic expression.

In the light of equation \eqref{Phi},
The electromagnetic field strength takes on the simple form
\begin{equation}
 F=\ell\wedge E+B,
 \label{modmaxF}
\end{equation}
where
\begin{equation}
 E=e^{-\gamma}\frac{g\sin\Phi-q\cos\Phi}{r^2+n^2}dr,
\end{equation}
and
\begin{equation}
 B=(q\sin\Phi+g\cos\Phi)\omega.
\end{equation}
A straightforward calculation shows that $P$ is given, in terms of $E$ and $B$, by
\begin{equation}
 P=\ell\wedge \left(e^{\gamma}E\right)+e^{-\gamma}B.
 \label{modmaxP}
\end{equation}

On the gravity side, we have Einstein's equations
\begin{equation}
R^\mu{}_\nu - \frac 12 \delta^\mu_\nu R +\Lambda \delta^\mu_\nu = 8\pi T^\mu{}_\nu,
\label{fieldequ}
\end{equation}
where the energy-momentum tensor is given by \eqref{Tmunu}; they imply that the metric function
is fixed by
\begin{equation}
f(r) = \frac{\kappa(r^2-n^2) -2mr  + e^{-\gamma}(q^2 + g^2) - \Lambda (r^4 + 6 n^2 r^2 - 3 n^4)/3.}{r^2+n^2}
\label{sol}
\end{equation}
It is clear that in the limit $n \to 0$, together with $\Lambda=0$ and $\kappa=1$, we recover the ModMax black hole solution discussed in~\cite{Flores-Alfonso:2020euz}.

In the following sections, we always consider a negative cosmological constant.
Under this restriction the configurations are asymptotically
locally AdS, i.e., in the asymptotic region the sectional curvature is constant (and negative) up to leading order.
The cosmological constant sets the AdS scale by $l^2=-3\Lambda^{-1}$.

\subsection{Traversable AdS Wormholes}
\label{secc:2a}

Taub-NUT geometries are stationary but not static and have arguably simpler metrics than other stationary spacetimes, e.g., Kerr solutions and the like. This allows for the further development of research widely applied to static configurations. Spontaneous scalarization is an example where the first (theoretical) observation of the phenomenon on non-static geometries was provided by a Taub-NUT Ansatz~\cite{Brihaye:2018bgc}. Another such case has been provided in~\cite{Corral:2019leh}, where Wheeler polynomials have been generalised to stationary spacetimes for the first time, by virtue of NUTty metrics.

Another consequence of the work done in Ref.~\cite{Brihaye:2018bgc} is that scalarised NUTty wormholes have been very recently found~\cite{Ibadov:2020ajr}; these wormholes are generalisations of those found in Ref.~\cite{Antoniou:2019awm}. In contrast, the traversable wormholes we construct in this section are sourced by self-interacting electromagnetic matter. To reach this goal, let us consider first a hyperbolically symmetric Taub-NUT metric in standard coordinates~\cite{Griffiths:2009dfa}
\begin{equation}
 ds^2 =-f(r)\left[dt+4n\sinh^2{(\psi/2)}d\phi\right]^2+f(r)^{-1}dr^2+(r^2+n^2)\left(d \psi^2+\sinh^2\psi d\phi^2\right).
\end{equation}
Now, let us use the hyperbolic identity $\sinh^2{(\psi/2)}=(\cosh\psi-1)/2$ and then define $u=t-2n\phi$ so that
\begin{equation}
 ds^2 =-f(r)\left[du+2n\cosh\psi d\phi\right]^2+f(r)^{-1}dr^2+(r^2+n^2)\left(d \psi^2+\sinh^2\psi d\phi^2\right).
\end{equation}
We now switch to Lobachevsky coordinates. For simplicity, let us recycle the coordinate $t$ and define it anew. Thus, we write the Lobachevsky coordinates as
\begin{equation}
 t=\arctanh(\tanh\psi\cos\phi), \quad\text{and}\quad
 \theta=\arcsinh(\sinh\psi\sin\phi),
\end{equation}
which in turn yield
\begin{equation}
 ds^2 =-f(r)\left[du+2n\sinh\theta d t\right]^2+f(r)^{-1}dr^2+(r^2+n^2)\left(\cosh^2\theta d t^2+d \theta^2\right). \label{WLP}
\end{equation}
This metric is stationary and axisymmetric; in particular, it can be written in the Weyl-Lewis-Papapetrou (WLP) from. It is well known, that there are
formal coordinate changes which map WLP metrics into
Einstein-Rosen metrics or to Gowdy metrics.
In any case, these are classes of metrics which are very closely related; see, e.g.~\cite{Hernandez:2008wm}. In particular, a double Wick rotation, $u\to\mbox{i}u$ and $t\to\mbox{i}t$, allows us to write a wormhole metric from \eqref{WLP} as
\begin{equation}
 ds^2 =f(r)\left[du+2n\sinh\theta dt\right]^2+f(r)^{-1}dr^2+(r^2+n^2)\left(-\cosh^2\theta dt^2+d \theta^2\right). \label{ER}
\end{equation}
Taking equation \eqref{gauge} with
\begin{equation}
 a(r)=-\frac{q\sin\Phi+p\cos\Phi}{2n}
\end{equation}
and $\ell=du+2n\sinh\theta dt$ results in a metric function
\begin{equation}
 f(r)=\frac{-r^2+n^2 -2mr  - e^{-\gamma}(q^2 + g^2) + l^{-2}(r^4 + 6 n^2 r^2 - 3 n^4)}{r^2+n^2}.
\end{equation}
This geometry is sourced by fields $F$ and $P$ of the form described in equations 
\eqref{modmaxF} through \eqref{modmaxP}.
After rescaling some coordinates and parameters, the vacuum solution of Ref.~\cite{Anabalon:2018rzq} is recovered when the fields are turned off. Moreover, the electrovacuum solution of Ref.~\cite{Anabalon:2020loe} is recovered in the linear limit, $\gamma=0$. In the special case $q=p=m=0$ and $n=l/2$ the metric is exactly AdS$_4$, cf. Sec. 2.4 of Ref. \cite{Johnson:2014xza}.

\begin{figure}[tbp]
\centering
\includegraphics[width=0.45\textwidth]{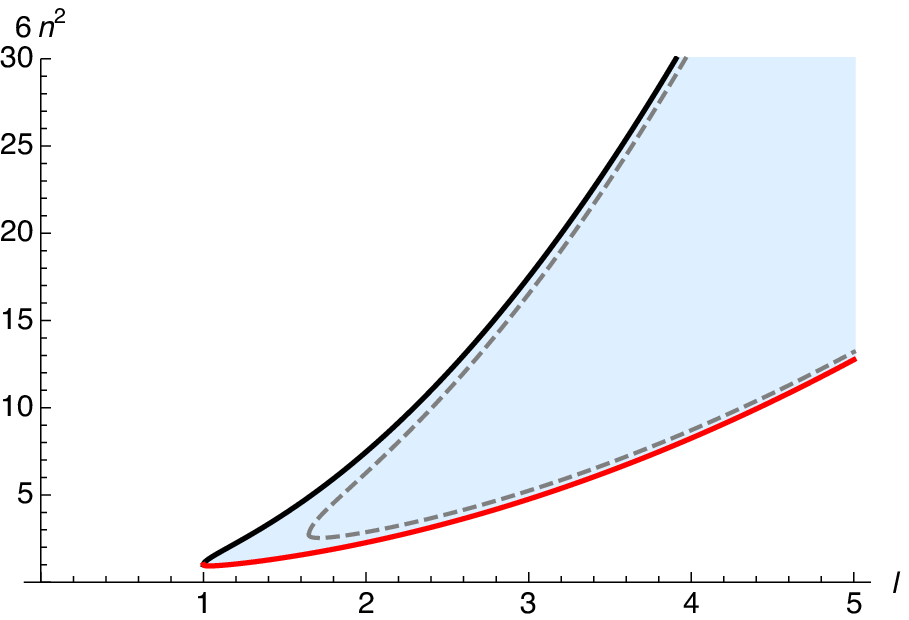}\hfill\includegraphics[width=0.45\textwidth]{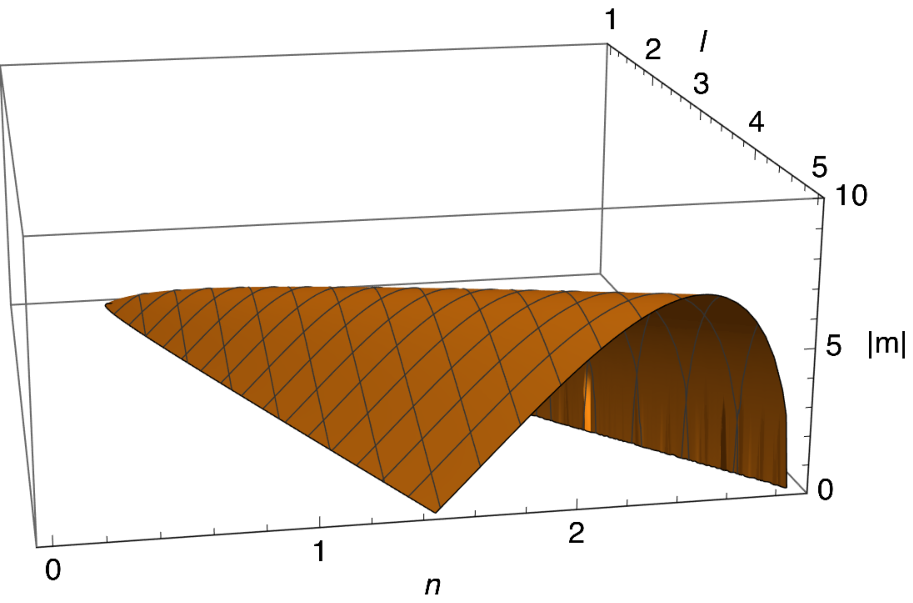}
\caption{Plots showing the restrictions in parameter space for the existence of a wormhole. The left panel shows in the $(l, 6n^2)$-plane, the region where the metric function $f(r)$ has no zeroes and is positive definite. The upper (black) curve corresponds to $6 n^2 = l^2 + l \sqrt{l^2-12 X}$ whilst the lower (red) curve to $6n^2 = l^2 - \frac 12 l \sqrt{l^2-12 X}$; the dotted (gray) curves correspond to the Maxwell case. In the right panel, only the region below the displayed surface corresponds to the permissible values for $|m|$. For illustrative purposes, we have set $X := e^{-\gamma} (q^2 + g^2) = 1/12$ and $\gamma = 1$ so that $X_{Maxwell} := q^2 + g^2 = e/12$.}
\label{inequalities}
\end{figure}

For the metric in Eq.~\eqref{ER} to represent a wormhole, the function $f(r)$ must be everywhere positive. It is straightforward to see that this condition is satisfied when
\begin{equation}
|m| < \frac{\sqrt{l^2 + l \sqrt{l^2-12 X} - 6 n^2} \left(12 n^2 + l \sqrt{l^2-12 X} - 2 l^2 \right)}{3 \sqrt{6} l^2},
\label{wormhole1}
\end{equation}
where $X := e^{-\gamma} (q^2 + g^2)$. Obviously, $12 X \leq l^2$ must hold together with $ 6 n^2 < l^2 + l \sqrt{l^2-12 X}$ and $2l^2 - l \sqrt{l^2-12 X}  < 12 n^2$; the last two inequalities imply
\begin{equation}
l^2 - \frac 12 l \sqrt{l^2-12 X} < 6 n^2 < l^2 + l \sqrt{l^2-12 X}.
\label{wormhole2}
\end{equation}
Save for the factor $e^{-\gamma}$ in the definition of $X$, these expressions are similar to those of the Maxwell scenario. We summarise and illustrate these restrictions in Fig.~\ref{inequalities}.

Notice that hypersurfaces defined by $r=\text{constant}$
have a squashed Coussaert-Henneaux geometry with
$\mathfrak{so}(2,1) \oplus \mathfrak{so}(2)$ isometry algebra~\cite{Coussaert:1994tu,AyonBeato:2004if}. Coussaert-Henneaux spacetime is a fibration over AdS$_2$; this part of the metric is responsible for the SO(2,1) isometry. Before the double Wick rotation, $r=\text{constant}$ submanifolds are fibrations over $H^2$ which share isometry group with AdS$_2$; this indirectly points to why this double Wick rotation procedure is particular to Taub-NUT spaces with topologically trivial base manifolds. For spherically symmetric
configurations this is impossible because 
$r=\text{constant}$ surfaces are Lorentzian hyperspheres; the bundle structure is that of Hopf.
Since the base sphere cannot support Lorentzian metrics then there is no analogue formal transformation for Taub-NUTs with spherical symmetry.

In what follows our attention shifts to Euclidean configurations. Hence, to close the part of our work dealing with Lorentzian geometries, we mention in what way the configurations presented above are new. They are new solutions specific to GR coupled to ModMax theory. The nonlinear constitutive relation is a one-parameter family generalising Maxwell's. Nonetheless, its equations of motion enjoy SO(2) duality invariance and so are a special case of the infinite class considered in~\cite{Salazar:1987ap}. Thus, the nonlinearly charged Taub-NUT systems we provide are a special case of those found there as well. They correspond to the unique case where the matter source is conformal. This includes the wormhole geometries. Notwithstanding, specialising to the conformal case has allowed for 
their non-trivial physical interpretation as traversable AdS wormholes.

\subsection{The NUTs and Bolts of ModMax}
\label{secc:2b}

In this section, we write down Euclidean geometries  which have the form \eqref{taubnut}. These spaces began to be studied because of their analogy with Yang-Mills instantons~\cite{Hawking:1976jb}. For recent work in this direction we suggest~\cite{Arratia:2020hoy}; therein scalar fields in Horndeski theory source the spaces. Our purpose here is to use these geometries as gravitational instantons to carry out semi-classical calculations. For this purpose the Taub-NUT configurations must be regular.

For us, in the classical action one varies with respect to the metric or the gauge potential. Useful calculations in Euclidean quantum gravity involve approximating the partition function using a saddle point approximation. In practice, one restricts oneself to spaces with some particular symmetry so that the integrals remain under control. Boundary conditions are fixed whilst all infilling geometries and topologies are considered. Only non-singular Euclidean configurations dominate the path integral, hence, these are the relevant geometries and potentials to consider.

The Euclidean configurations are readily found by the Wick rotation $t\to \mbox{i}\tau$, together with the corresponding changes
$n\to \mbox{i}n$ and $q\to \mbox{i}q$. In other words, the positive definite metric is
\begin{equation}
ds^2 = \frac{V(r)}{r^2-n^2}(d\tau + 2n{\cal A})^2 +\frac{r^2-n^2}{V(r)}  dr^2 +  (r^2 - n^2)d\sigma^2,
\label{etaubnut}
\end{equation}
where
\begin{equation}
 V(r)=\kappa(r^2+n^2) -2mr  + e^{-\gamma}(g^2 - q^2) - \Lambda (r^4 - 6 n^2 r^2 - 3 n^4)/3.
 \label{eV}
\end{equation}
For each base manifold, there are corresponding symmetry restrictions and boundary conditions, e.g., when $\kappa=1$ the metric has isometry group SU(2)$\times$U(1). In this case, all possible infilling geometries must possess this isometry. Also, every possible topology has $S^3$ as its asymptotic boundary.

The Euclidean gauge potential consistently incorporates the same changes as the metric, i.e.,
\begin{equation}
 A = \frac {1}{2n} (-q \sinh \varPhi + g \cosh \varPhi) \ell, \label{egauge}
\end{equation}
where $\ell=d\tau + 2n{\cal A}$ and
\begin{equation}
 \varPhi(r)=2n e^{-\gamma} \int_r^\infty \frac {dr}{r^2 - n^2} = 2 e^{-\gamma} \arccoth\left( \frac r n \right).
 \label{ePhi}
\end{equation}
The electromagnetic fields, $F$ and $P$, still have the forms of \eqref{modmaxF} and \eqref{modmaxP}, respectively. However, we now have
\begin{equation}
 E=e^{-\gamma}\frac{q\cosh\varPhi-g\sinh\varPhi}{r^2-n^2}dr,
\end{equation}
and
\begin{equation}
 B=(g\cosh\varPhi-q\sinh\varPhi)\omega.
\end{equation}

These configurations must be free of any type of singularity. As they come from Lorentzian metrics which posses Killing horizons then we examine the fixed point set of $\xi=\partial/\partial \tau$. These come in two forms: 1) zero dimensional sets called \emph{nuts} which only happen when $r=n$. 2) codimension two sets dubbed \emph{bolts} which must happen at some value $r=r_b>n$. Following standard parlance, we refer to solutions with a bolt as Taub-Bolt.

Let us focus first on configurations with a nut.
We start by analysing the gauge potential. Returning to our notation $A=a\ell$ allows us to note that the potential is regular at $r=n$ if $a(n)$ is regular. However, $\ell$ is not defined here, thus,
for $A$ to be regular we must enforce that $a(n)=0$. This only happens when $q=g$. As a consequence the energy momentum tensor vanishes. This condition is reflected in the metric by the vanishing of the term $e^{-\gamma}(g^2 - q^2)$ in the metric function \eqref{eV}. Thus, the electromagnetic matter is stealthy, in the sense of \cite{AyonBeato:2004ig}. This is a general phenomenon for selfdual Euclidean Maxwell fields; see for example \cite{Aliev:2005pw} and references therein. For Taub-NUT spaces there is also a relation between selfduality, or the lack thereof, and topological charge~\cite{Flores-Alfonso:2018jra}. Because of the topological nature of this result, it also holds for nonlinear electrodynamics; see e.g.~\cite{Flores-Alfonso:2018aaw}.

In nonlinear electrodynamics, selfduality is not a concept that applies to fields as the Hodge map is linear while $F$ and $P$ are by definition related in a nonlinear manner. Nonetheless, in a theory with electric-magnetic duality invariance
(anit-)selfduality means that electric and magnetic charges are equal (opposite resp.).
For configurations with a nut, the regularity condition $q^2=g^2$
is an imposition of (anti-)selfduality,
\begin{equation}
 \int F=\pm \int *P,
\end{equation}
where we have denoted the Hodge star operation by $*$ to emphasise it is Euclidean in this instance.

For the gauge potential \eqref{egauge} to be regular the charges must satisfy $q=g$ which leads to fields
\begin{equation}
 F=ge^{-\varPhi}\left(\frac{e^{-\gamma}}{r^2-n^2}\ell\wedge dr+\omega\right),
\end{equation}
and
\begin{equation}
 P=ge^{-\varPhi}\left(\frac{1}{r^2-n^2}\ell\wedge dr+e^{-\gamma}\omega\right).
\end{equation}
The stealthy behaviour of these fields is comparable to the Maxwell case and is a newly explored feature of ModMax electrodynamics.

Since, the metric is now equivalent to a vacuum solution then regularisation of the metric is carried out as in \cite{Chamblin:1998pz},
yielding
\begin{equation}
 V_n(r)=\frac{r-n}{r+n}\left[\kappa+l^{-2}(r-n)(r+3n)\right],
\end{equation}
only for $\kappa=0,1$; hyperbolic configurations are not consistent with nut-type fixed points of $\xi$. For vanishing cosmological constant, planar configurations do not exist. Under such conditions, we recover the original Euclidean Taub-NUT metric of~\cite{Hawking:1976jb}.

We emphasise that, for both Taub-NUT and Taub-Bolt systems, only the spherical case
requires regularisation due to conical singularities; these come from the Euclidean analogue of the Misner string and require $\tau\sim\tau+8\pi n$. For non-spherical configurations
the Euclidean time $\tau$ admits arbitrary periodicities.

We now turn to Taub-Bolt configurations. Firstly, we note that
the Euclidean electric potential at infinity, $v$, is given by
\begin{equation}
 v=\frac{g}{2n}.
\end{equation}
For ease of comparison, we have parametrised our configurations in terms of the value of $v$, as was done first in~\cite{Awad:2005ff} and later in \cite{Johnson:2014xza,Johnson:2014pwa}.
Secondly, the condition $a(r_b)=0$ must be enforced so that $A$ is regular at the bolt, where $d\tau$ is not defined. Consequently, we have that
\begin{equation}
q_b = 2nv \coth \varPhi(r_b), \label{qb}
\end{equation}
where we have used the subscript $b$, as above,
to reference configurations with a bolt. Notice that the regularity condition implies setting the 
Euclidean electric potential at the bolt to zero. Thus, $v$ coincides with the \emph{potential difference} between infinity and the bolt.
With equation \eqref{qb} in mind, the definition of $r_b$, i.e., $V(r_b)=0$, yields an expression restricting the mass
\begin{equation}
m_b = \frac {\kappa(r_b^2 + n^2) - 4n^2v^2 e^{-\gamma} \csch^2 \Phi(r_b)}{2r_b} + \frac 1{2l^2} \left( r_b^3 - 6n^2 r_b - 3 \frac {n^4}{r_b} \right).
\end{equation}
Lastly, for the metric to be regular at the bolt then no conical singularities should arise there. As a consequence, the Euclidean time coordinate has periodicity $\Delta \tau = 4\pi/f_{,r}(r_b)$, where $f(r)=V(r)/(r^2-n^2).$

\begin{figure}[tbp]
\centering
\includegraphics[width=0.6\textwidth]{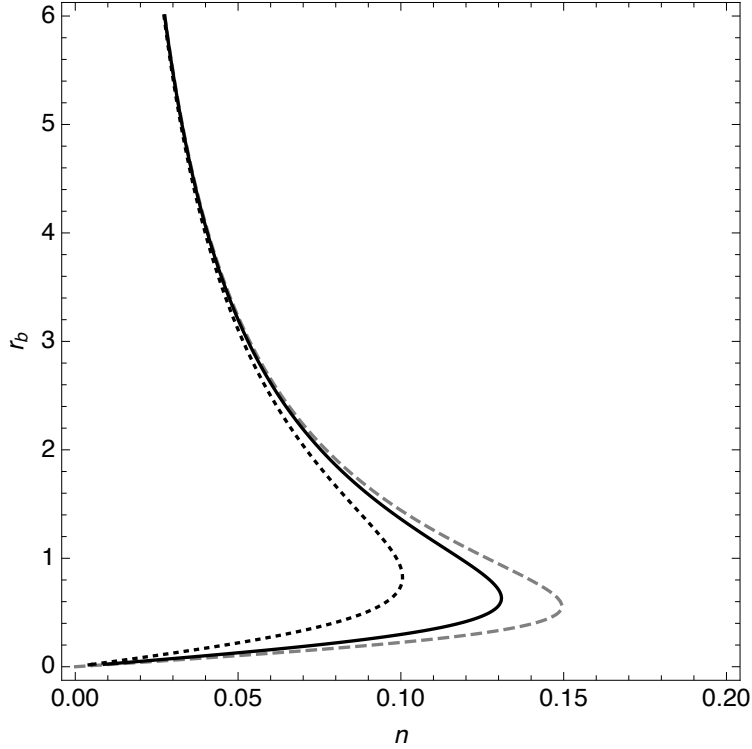}
\caption{Plot of the bolt radius $r_b$ as a function of $n$ with non-linear electrodynamics parameter $\gamma = 0.1$. The (grey) dashed line corresponds to the standard AdS-Taub-bolt solution with $v=0$ and $n_{max}^{AdSTb} = 0.149 l$, whilst the (black) solid and dotted lines correspond to $v = 1/2$ and $v = 1$ respectively; in these two cases, the maximum allowed value for $n$ is always smaller than $n_{max}^{AdSTb}$.}
\label{fig1}
\end{figure}

When spherically symmetric configurations are considered, 
$\kappa=1$, then an additional restriction needs to be considered. A conical singularity coming from the Misner string in the Lorentzian metric forces the identification $\Delta \tau=8\pi n$. This, in turn, allows us to write
\begin{equation}
6n r_b^2 - l^2 r_b - 6n^3 + 2n l^2 + 8 n^3 l^2 v^2 e^{-\gamma} \frac {\csch^2 \varPhi(r_b)}{r_b^2 - n^2 } = 0. \label{tras}
\end{equation}
Since $\varPhi(r_b) = 2 e^{-\gamma} \arccoth (r_b/n)$, we have a transcendental equation for $r_b$, cf. equation (31) of \cite{Johnson:2014pwa} which is algebraic. Indeed, in the linear case ($\gamma=0$) equation \eqref{tras} becomes algebraic and coincides with the Maxwell case, as it should. In Fig.~\ref{fig1}, we illustrate the behaviour of $r_b(n)$ for the choices $v = 1/2, 1$ and we compare it with the standard AdS-Taub-Bolt solution~\cite{Johnson:2014xza}. The relevance of this plot is that $n$ is in one-to-one correspondence with the temperature of the system.
A maximum value of $n$ indicates a minimum temperature for which bolt configurations may exist.
We see that the maximum value of $n$ in our configurations decreases as $v$ increases whilst keeping the non-linear parameter $\gamma$ fixed. The two branches of $r_b$ for small values of $n$ behave as
\begin{equation}
r_{b-} = \left( 1 + v^2 e^\gamma \right) 2n + O(n^3), \qquad \text{and}  \qquad r_{b+} = \frac {l^2}{6n} - \left( 1 + v^2 e^\gamma \right) 2n + O(n^3)
\label{nexpansion}
\end{equation}
showing that the small branch smoothly connects to the vacuum case $r_b=2n$. This is shown graphically in Fig.~\ref{fig1}. Moreover, these expressions indicate that the two branches join smoothly. All these features are similar to those present in the  Maxwell case~\cite{Johnson:2014pwa}.

Notice that for arbitrarily large but negative values of $\gamma$ \eqref{tras} becomes identical to the neutral case. This is also visible in equations \eqref{nexpansion}.
The nonlinearity parameter, $\gamma$, smoothly connects the neutral and Maxwell cases whilst maintaining the qualitative behavior of the system.

\section{Thermodynamic Phase Structure}
\label{secc:3}

In this section, we explore the thermodynamic behavior of the spherical ($\kappa=1$) Taub-NUT/Bolt system. Our main goal is to obtain the Euclidean on-shell action, as it is in correspondence with the system's free energy. A lower value of the action indicates which thermodynamic state is preferred by the system. 
Varying only the nut parameter tells us which phases dominate the path integral at high or low temperatures. Let us keep in mind how the temperature $T$, the Euclidean time period $\beta$ and the nut parameter $n$ are all related
\begin{equation}
 T=\frac{1}{\beta}=\frac{1}{8\pi n}.
\end{equation}

We consider the standard framework of the black hole thermodynamics~\cite{Bardeen:1973gs,Bekenstein:1973ur,Hawking:1974sw,Hawking:1976de,Gibbons:1976ue} and apply it to Euclidean geometries with nuts or bolts, as in~\cite{Chamblin:1998pz,Arratia:2020hoy}.  
Let us recall that our boundary conditions fix $\beta$, and so $T$ and $n$ with it. They also fix the gauge potential at infinity, $v$. As a consequence, the system's free energy has control variables $T$ and $v$. In what follows, our main goal is to explore the effect the ModMax nonlinear extension has on the system. Hence, comparison with reference~\cite{Johnson:2014pwa} is key.

The conformal nature of the nonlinear electromagnetic matter source leads to
\begin{equation}
 R=4\Lambda,
\end{equation}
as in the Maxwell case. As a consequence, calculating the on-shell Euclidean action, $I$, in a standard way~\cite{Emparan:1999pm} is equivalent. The main difference is the contribution of the matter Lagrangian which can be calculated from $F$ and $P$ through
\begin{equation}
L =-\frac 14 F_{\mu\nu}P^{\mu\nu}= -\frac 12 \frac {(q^2 + g^2)\cosh 2\varPhi - 2 qg\sinh 2\varPhi}{(r^2 - n^2)^2} e^{- \gamma} ,
\end{equation}
from which it follows that
\begin{equation}
I= 4\pi n \left[ m + \frac {3 n^2 r_+ - r_+^3}{l^2} +\frac{g^2}{2n}\coth\varPhi(r_+) \right].
\end{equation}

Notice that, in our notation, the action of the nut case is
\begin{equation}
 I_n=4\pi n^2\left(1-\frac{2n^2}{l^2}+2v^2\right),
\end{equation}
which appears to be the same as the Maxwell case. However, let us recall that $v=g/2n$ and that $g$ is obtained from $*P$ and not $*F$. In ModMax theory the distinction between the two involves a factor of $e^{-\gamma}$; see equations \eqref{modmaxF} and \eqref{modmaxP}. In this sense, the only difference between the action presented here and the linear case is an exponential of $\gamma$ in the electromagnetic contribution.

Let us approach this issue from a different perspective to further emphasise the distinction between self-dual fields in Maxwell theory and self-dual configurations in nonlinear electrodynamics.

To begin with, notice that for spacetimes with a nut, the field strength does not comply with $F=*F$, however, it does satisfy $F=*P$. In particular, this implies that the matter action is proportional to the Chern-Pontryagin index, see, e.g., Ref.~\cite{Jackiw:2005wp}. Symbolically,
\begin{equation}
 -\frac{1}{4}\int d^4x\sqrt{g}F_{\mu\nu}P^{\mu\nu}
 =-\frac{1}{2}\int F\wedge *P=-\frac{1}{2}\int F\wedge F.
\end{equation}
Moreover, by examining Eqs.~\eqref{egauge} and \eqref{ePhi} we see that the Euclidean ModMax configuration is asymptotically Maxwell. This is to say, the results of Refs.~\cite{Flores-Alfonso:2018jra,Flores-Alfonso:2018aaw} apply
\begin{equation}
 \frac{1}{4\pi^2}\int F\wedge F=4g^2.
\end{equation}
Therefore, there is no modification induced by the non-linear parameter $\gamma$ on this topological charge.

Furthermore, as mentioned above, we are working in a setup with fixed temperature and fixed potential. The Euclidean action yields the free energy by $W=I/\beta=U-TS-vQ$~\cite{Chamblin:1999tk}. Hence, it follows that
\begin{align}
 U&=\left(\frac{\partial I_n}{\partial \beta}\right)_{v}-\frac{v}{\beta}\left(\frac{\partial I_n}{\partial v}\right)_{\beta}=n-\frac{4n^3}{l^2}=m
 \label{U}\\
 S&=\beta\left(\frac{\partial I_n}{\partial \beta}\right)_{v}-I_n=4\pi n^2\left(1-\frac{6n^2}{l^2}+2v^2\right)\\
 Q&=-\frac{1}{\beta}\left(\frac{\partial I_n}{\partial v}\right)_{\beta}=-2nv=-q \label{Q}
\end{align}
Notice that Eq. \eqref{Q} is consistent with our notation in \eqref{egauge}, which is a consequence of how we carried out the Wick rotation. Together Eqs. \eqref{U}-\eqref{Q} satisfy
\begin{equation}
 d U=TdS+vdQ.
\end{equation}

\begin{figure}[tbp]
\centering
\includegraphics[width=.45\textwidth]{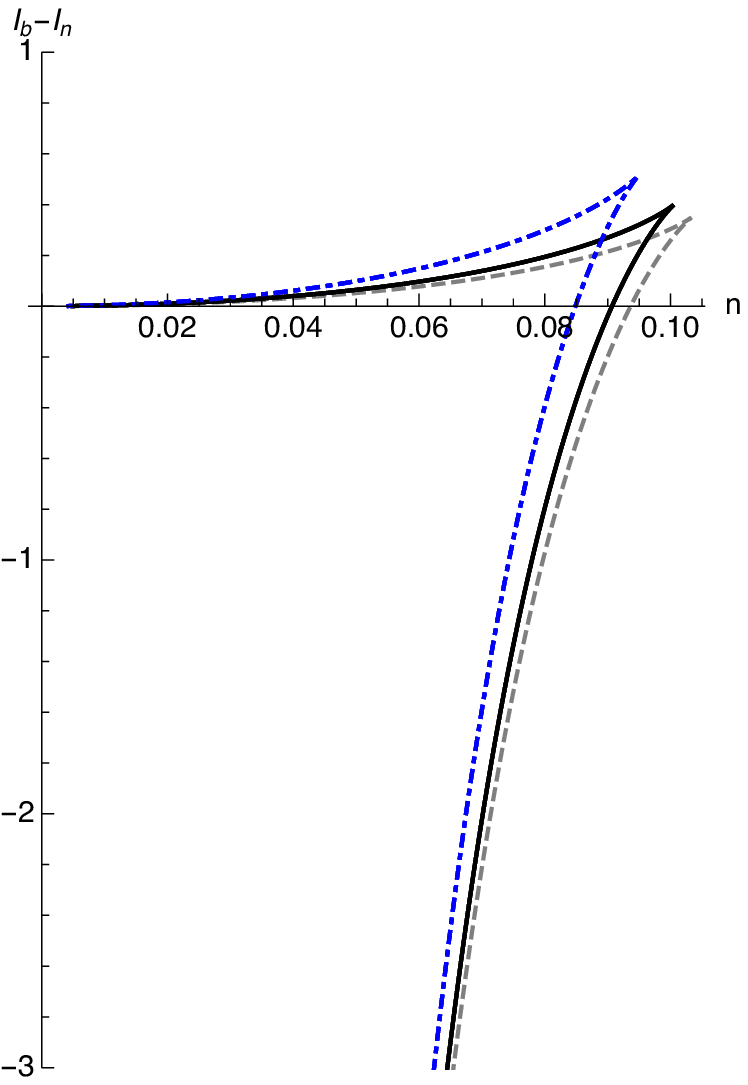}
\hfill
\includegraphics[width=.45\textwidth]{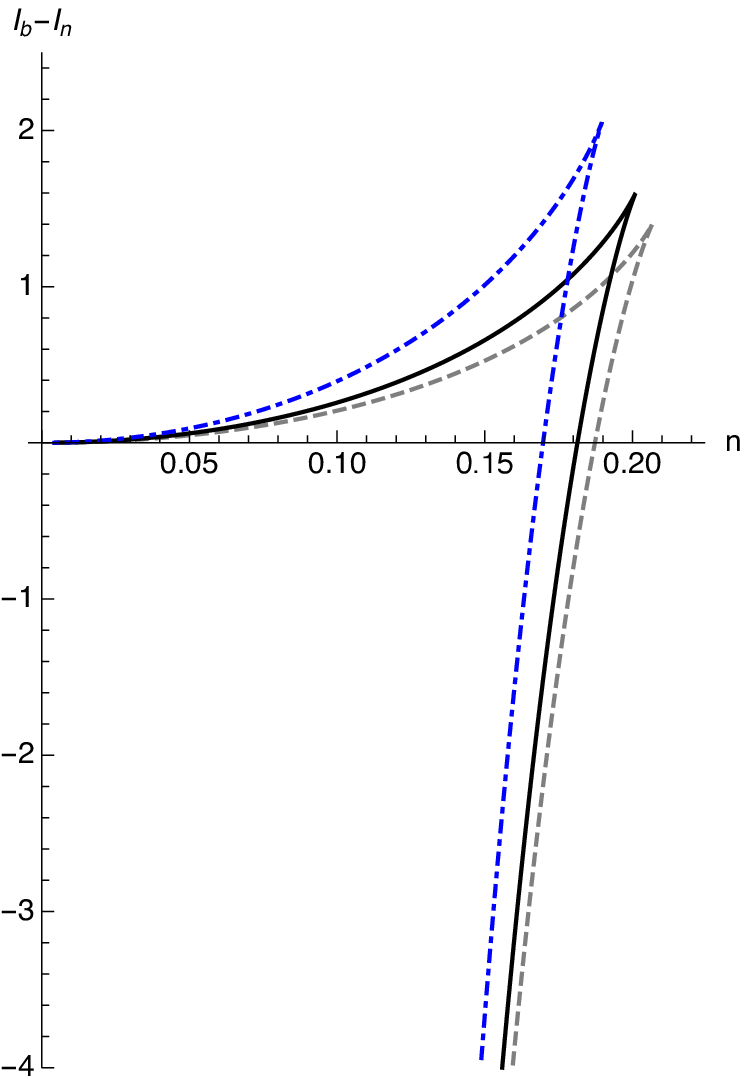}
\caption{Plot of the action difference $I_b - I_n$ for the AdS-Taub-NUT ModMax spacetime. The left panel corresponds to $l=1$, whilst the right panel to $l=2$. From right to left, we have the values $\gamma = 0, 0.1, 0.3$ (dashed, solid and dot-dashed lines respectively). As $\gamma$ increases, the curves move to the right so that both the critical temperature and the cusp temperature increase for a given value of $l$.}
\label{fig2}
\end{figure}

Now, motivated by the well-known phase structure of neutral and linearly charged Taub-NUT/Bolt systems, we examine the on-shell actions of the cases at hand. Let us keep in mind that, thermodynamically favoured configurations have a lower value of the action, as it is in correspondence with the system's free energy. For the Taub-NUT/Taub-Bolt AdS-ModMax spacetimes the difference between the corresponding actions is plotted in Fig.~\ref{fig2}, where we use values $l=1, 2$ and $v=1$. As in the standard Maxwellian case, there is a first order phase transition associated between nuts and large bolts. The temperature $(8\pi n_c)^{-1}$, associated to the value $n = n_c$ where the cusp shape appears, takes higher values as the non-linear electrodynamics parameter $\gamma$ increases for a given value of $l$. At the same time, the critical  temperature where the first order transition takes place, also increases with $\gamma$.

\section{Conclusions}

In this paper, we consider the nonlinear electrodynamics recently proposed in~\cite{Bandos:2020jsw}, referred to as ModMax. By considering standard gravity,
we have constructed various new self-gravitating electromagnetic fields. The fields self-interact through a nonlinear constitutive relation. They correspond to nonlinear extensions of charged Taub-NUT-type configurations. Turning off the NUT parameter yields topological black holes sourced by ModMax fields whose asymptotic behavior depends on the cosmological constant.

Among this family we count two non-diffeomorphic hyperbolically symmetric solutions. Both are radial extensions of fiber bundles over a base manifold with SO(2,1) symmetry. However, one of these base spaces is a Lobachevsky plane (Euclidean) and the other is the Anti-de Sitter plane (Lorentzian). The latter solution represents a nonlinearly charged traversable AdS wormhole whenever the metric function is positive everywhere,
cf. Eqs. \eqref{wormhole1} and \eqref{wormhole2} along with Fig.~\ref{inequalities}. Notice that for $\gamma>0$ a screening effect takes place. Thus, a comparison between Maxwell and ModMax configurations with the same physical parameters shows that due to charge screening, a larger area of parameter space is available for wormhole spacetimes to exist in the latter case; this difference is portrayed in Fig.~\ref{inequalities}.

Moreover, we emphasise that this type of non-equivalent pair solutions does not exist when spherical symmetry is considered as two-dimensional spheres do not support Lorentzian metrics.

Since ModMax is a nonlinear theory with SO(2) duality invariant motion then the spacetimes we have constructed belong to the infinite class of Taub-NUT configurations studied in~\cite{Salazar:1987ap}. By specialising to ModMax, the fields and every aspect of the geometries in that work can be fully specified and correspond to our results.

These configurations naturally lead to gravitational instantons which allow for thermodynamic explorations of non-black hole spacetimes. For such a purpose we specialise to a negative cosmological constant. We present new gravitating Euclidean nonlinear fields and analyse their regularity conditions. Configurations with a nut-type fixed point of the Euclidean time direction are self-dual and are supported by a background which is topologically trivial.
This self-duality makes the fields 
stealthy, i.e., their energy momentum tensor vanishes cf.~\cite{AyonBeato:2004ig}. Another effect of imposing regularity is that hyperbolically symmetric spaces are excluded. Moreover, spherically symmetric configurations are the only ones to smoothly
connect with the case of vanishing cosmological constant. In contrast, Taub-Bolt systems lack self-duality and generate backgrounds of nontrivial homology. They also admit any type of symmetry and smoothly connect to the case $\Lambda=0$.

Specialising to spherically symmetric instantons, we find nut/bolt phase transitions \`a la Hawking-Page which are comparable to the Maxwell case, cf. Fig. \ref{fig2}. In ModMax theory nonlinearity is measured by $\gamma$, which we show acts as a deformation parameter in phase space.

\acknowledgments
The authors thank Eloy Ay\'on-Beato and 
Crist\'obal Corral for insightful comments. We thank an anonymous referee for suggesting some clarifications. DFA is supported by a CONACYT postdoctoral fellowship. This work has been partially funded by Grant No. A1-S-11548 from CONACYT.

\paragraph*{\bf{Note added.}} After completing most of the work described in this manuscript, Ref.~\cite{Bordo:2020aoz} appeared, in which the solution of Sec. \ref{secc:2}, cf. \eqref{sol}, is displayed, albeit restricted to the case $\kappa=1$ and $\Lambda<0$. That paper also notes that $\gamma=0$ corresponds to the dyonic black hole of Ref.~\cite{Flores-Alfonso:2020euz}.

\end{document}